\documentclass[letterpaper, 10 pt, conference]{ieeeconf}  

\IEEEoverridecommandlockouts                              

\usepackage{amsmath} 
\usepackage{graphicx}

\title{\LARGE \bf
Poxel: Voxel Reconstruction for 3D Printing
}

\author{Ruixiang Cao$^{1}$ Satoshi Yagi$^{1}$ Satoshi Yamamori$^{2}$ and Jun Morimoto$^{1,2}$
\thanks{$^{1}$Graduate School of Informatics, Kyoto University, Kyoto, Japan
        {\tt\small cao.ruixiang.65h@st.kyoto-u.ac.jp, \{yagi, morimoto\}@i.kyoto-u.ac.jp}}
\thanks{$^{2}$Dept. of Brain Robot Interface, Computational Neuroscience Labs, ATR, Kyoto, Japan
        {\tt\small yamamori@atr.jp}
        }
        }

\begin{document}

\maketitle
\thispagestyle{empty}
\pagestyle{empty}

\begin{abstract}

Recent advancements in 3D reconstruction, especially through neural rendering approaches like Neural Radiance Fields (NeRF) and Plenoxel, have led to high-quality 3D visualizations. However, these methods are optimized for digital environments and employ view-dependent color models (RGB) and 2D splatting techniques, which do not translate well to physical 3D printing. This paper introduces "Poxel", which stands for Printable-Voxel, a voxel-based 3D reconstruction framework optimized for photopolymer jetting 3D printing, which allows for high-resolution, full-color 3D models using a CMYKWCl color model. Our framework directly outputs printable voxel grids by removing view-dependency and converting the digital RGB color space to a physical CMYKWCl color space suitable for multi-material jetting. The proposed system achieves better fidelity and quality in printed models, aligning with the requirements of physical 3D objects.

\end{abstract}

\IEEEpeerreviewmaketitle

\section{Introduction}

3D modeling and rendering have undergone substantial advancements over the past decade, enabling virtual 3D representations with exceptional realism. The progression from basic 2D imagery to high-dimensional, dynamic 3D reconstructions has been propelled by methods like Neural Radiance Fields (NeRF) \cite{mildenhall2020nerf} and Plenoxel \cite{plenoxel2021}. These approaches have yielded state-of-the-art results in generating 3D models optimized for digital visualization, such as VR and AR applications. Despite their success in creating view-dependent 3D visuals, these approaches face challenges in adapting to the requirements of physical 3D printing.

Our work seeks to address this gap by introducing a new method, "Poxel," that focuses on generating printable 3D voxel models compatible with photopolymer jetting printers. Unlike traditional voxel reconstruction methods optimized for virtual media, Poxel eliminates view-dependence and uses a CMYKWCl color model compatible with physical printing. This enables accurate, high-resolution, full-color 3D reconstructions that align well with the unique constraints and capabilities of multi-material jetting photopolymer 3D printing technologies.

\begin{figure}[htbp]
    \centering
    \includegraphics[width=\linewidth]{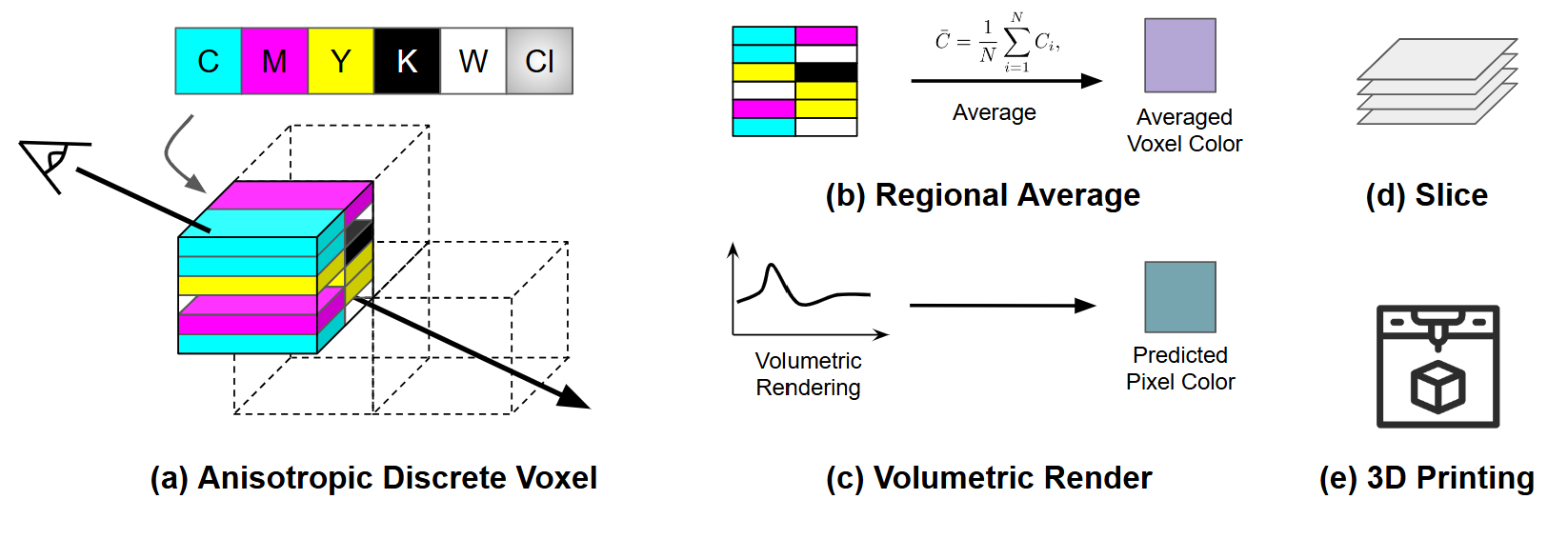} 
    \caption{Overview of the Poxel Framework for 3D Printing. (a) Anisotropic Discrete Voxel: A voxel structure with layered CMYKWCl (Cyan, Magenta, Yellow, Black, White, Clear) inks to capture color properties. The dimension of each voxel is anisotropic. (b) Regional Average: Averaging colors within voxel regions to achieve wide range of colors. (c) Volumetric Render: Predicts pixel color by integrating voxel color information. (d) Slice: Converts voxel data into printable layers for 3D printing. (e) 3D Printing: The final fabrication.}
    \label{overview}
\end{figure}

\begin{figure}[t]
    \centering
    \includegraphics[width=\linewidth]{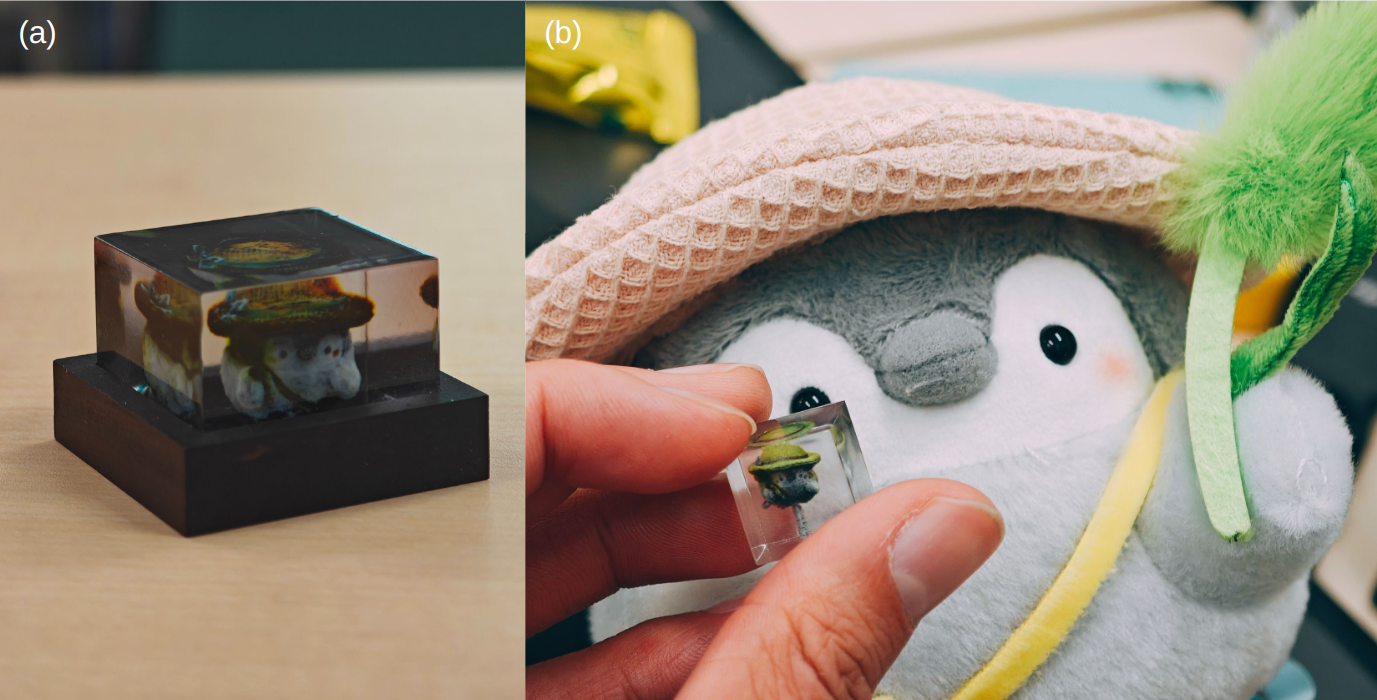} 
    \caption{The real-world experiment of Poxel. The model is printed by a Stratasys J850 printer
    (a) First attempt of the real-world test, with dimension of 30x30x15mm
    (b) Second attempt of the real-world test, with dimension of 15x15x15mm
    }
    \label{real}
\end{figure}

\section{Related Work}

\subsection{Neural Radiance Field}

NeRF \cite{mildenhall2020nerf} has been foundational in enabling view synthesis for complex 3D scenes. NeRF represents scenes as a volumetric function in a continuous domain, allowing high-quality renderings of novel views with impressive photorealism. However, NeRF relies on RGB color channels and view-dependent properties that are unsuitable for 3D printing. Its complexity in rendering detailed textures further complicates its application in generating physical models.

\subsection{Plenoxel}

Plenoxel \cite{plenoxel2021} builds on NeRF by replacing the implicit representation with an explicit voxel grid, making optimization more efficient. While this modification enables better control over the reconstructed 3D shapes, Plenoxel continues to depend on RGB colors and view-dependent rendering, limiting its applicability for 3D printing. Additionally, the method’s use of tri-linear interpolation leads to artifacts in physical models, where color accuracy and consistency are paramount.

\subsection{3D Gaussian Splatting}

3D Gaussian Splatting (3DGS) \cite{3dgs2023} achieves state-of-the-art results in 3D reconstruction for digital media by representing surfaces as 2D Gaussian splats optimized for view-dependent rendering. This approach suits VR and gaming environments but fails to provide view-independent, printable structures. Furthermore, its reliance on 2D splats rather than a volumetric representation complicates adaptation to the physical requirements of 3D printing.

\subsection{Multi-Material Jetting Photopolymer 3D Printing}

Multi-material jetting technology \cite{stratasys, sailner, mimaki} uses photopolymers and UV inks, offering full-color, high-resolution 3D printing by mixing CMYKWCl inks. This method allows for resolutions up to 600x300 dpi with layer thicknesses as small as 0.014mm, making it suitable for creating detailed and vibrant models. Such printers can reproduce fine color gradients and high detail, but require precise CMYKWCl-based color control and a voxel-based representation tailored to physical media, both of which Poxel addresses.

\section{Method}

In this work, we introduce a novel voxel-based reconstruction method specifically optimized for 3D printing, addressing limitations present in methods designed for digital visualization, such as Neural Radiance Fields (NeRF) \cite{mildenhall2020nerf}, Plenoxel \cite{plenoxel2021}, and 3D Gaussian Splatting (3DGS) \cite{3dgs2023}. Our approach, named "Poxel," generates CMYKWCl color-encoded voxels for high-fidelity, full-color 3D prints by eliminating view-dependency and focusing on color discretization suitable for physical media.

\subsection{Volume Rendering}

Traditional 3D rendering, such as NeRF, represents scenes as continuous volumetric functions mapping coordinates to RGB values, which work well for digital view-dependent applications but are unsuitable for static 3D printing \cite{mildenhall2020nerf}. Plenoxel enhances efficiency with explicit voxel grids, but RGB-based rendering and interpolation artifacts limit its applicability for physical models \cite{plenoxel2021}.

Our approach adapts these principles by using view-independent, discrete CMYKWCl voxel values, optimized for 3D printing. This configuration enables accurate color reproduction and stable structures suitable for physical prints.

\subsection{Anisotropic Voxel}

To improve print fidelity, we employ anisotropic voxels, unlike isotropic methods like 3D Gaussian Splatting that rely on 2D splats and view-dependent color \cite{3dgs2023}. Anisotropic voxels enable finer control over layer resolution, color blending, and voxel placement, minimizing artifacts and aligning with the precision needs of multi-material 3D printers.

\subsection{Color Discretization}

To ensure accurate color reproduction for 3D printing, our method discretizes RGB colors into CMYKWCl, tailored for multi-material jetting printers. Unlike continuous RGB gradients, which can cause artifacts, this discrete mapping produces stable, reproducible colors in physical models.

The discretization pipeline uses efficient, parallelized processing to manage the high voxel count needed for high-resolution prints. This ensures rapid color transformation while maintaining fidelity within the printer's color gamut, resulting in vibrant, consistent colors across the model.

\begin{figure}[t]
    \centering
    \includegraphics[width=\linewidth]{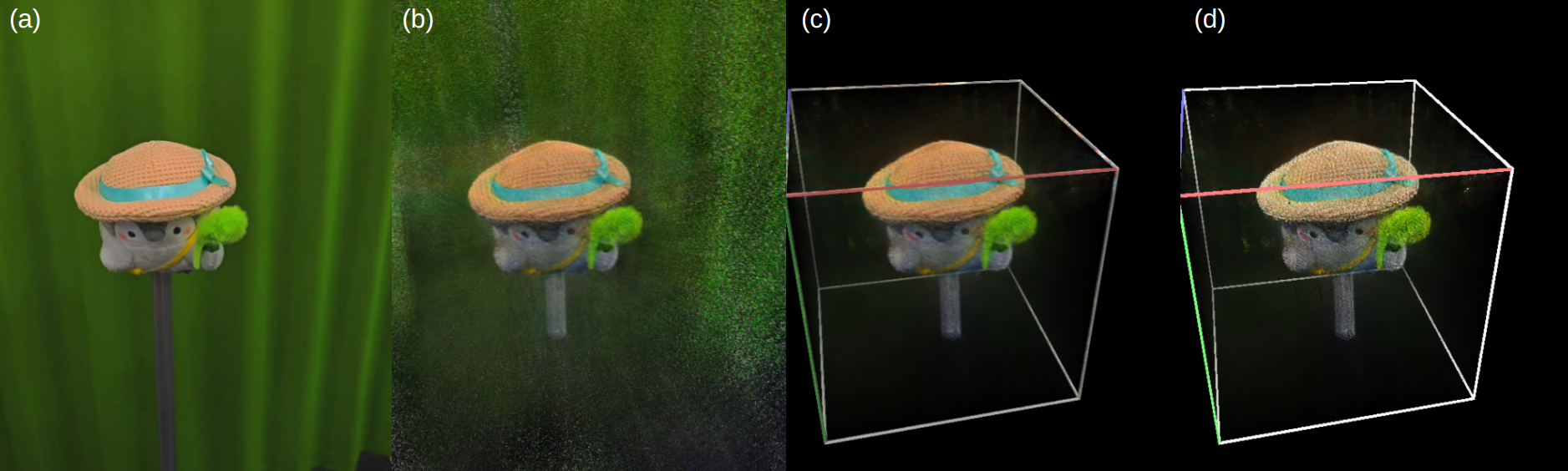} 
    \caption{End-to-end reconstruction result of Poxel before discretization (Third attempt).
    (a) Input figure taken from an iphone 14 smartphone. 
    (b) Reconstructed voxel grid after color averaging together with the background. 
    (c) Reconstructed voxel grid after color averaging without the background.
    (d) Reconstructed voxel grid before color averaging}
    \label{doll}
\end{figure}

\begin{figure*}[t]
    \centering
    \includegraphics[width=\linewidth]{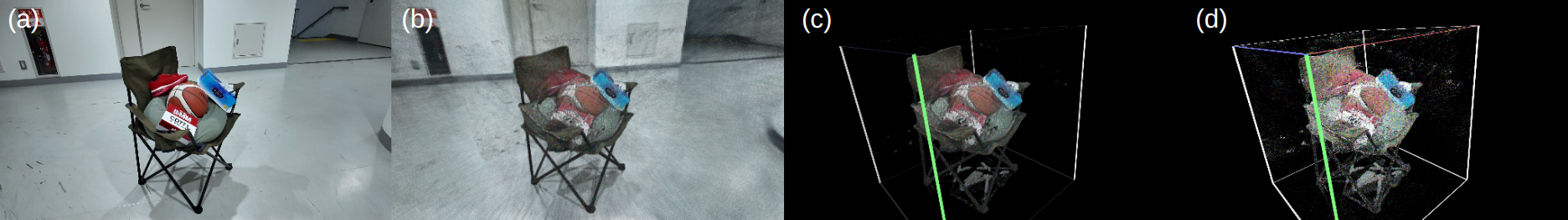} 
    \caption{End-to-end reconstruction result of Poxel 
    (a) Input figure taken from an DJI Osmo Pocket 3 camera\cite{dji_osmo_pocket_3}. 
    (b) Reconstructed voxel grid after color averaging together with the background. 
    (c) Reconstructed voxel grid after color averaging without the background.
    (d) Reconstructed voxel grid before color averaging}
    \label{greenchair}
\end{figure*}

\subsection{Optimization}

Our optimization strategy enhances color fidelity and structural alignment through loss functions focused on discrete color matching and spatial coherence. Instead of assigning colors directly to individual voxels, we average colors across a region (1x6x2) to achieve smoother transitions and reduce abrupt color changes.

The color mixture function averages RGBA values within this region as follows:
\begin{equation}
\bar{C} = \frac{1}{N} \sum_{i=1}^{N} C_i,
\end{equation}
where \( \bar{C} \) is the averaged RGBA color, \( C_i \) represents each voxel's color, and \( N = 12 \) for this region.

Structural loss encourages spatial consistency within voxel neighborhoods, fostering a cohesive appearance across complex geometries. Together, these loss functions refine color accuracy and alignment, ensuring high-quality, print-ready models with smooth color gradients and structural fidelity.

\section{Results}
We evaluated Poxel’s effectiveness through 3D printing experiments on the Stratasys J850, a full-color, multi-material printer supporting CMYKWCl inks. Two batches of printed models allowed comparison between traditional 3D Gaussian Splatting (3DGS) and our Poxel method. The model is trained with a single Nvidia RTX4090 GPU\cite{nvidia_rtx4090}.

In the first test\ref{real}, we adapted 3DGS for printing by converting 2D splats to 3D ellipsoidal volumes, slicing along the height axis, and mapping RGB values to a CMYKWCl palette. This process introduced artifacts and reduced fidelity due to additional conversion steps.

In the second test\ref{real}, Poxel’s end-to-end approach directly generated a CMYKWCl voxel grid optimized for printing, eliminating intermediate conversions. The loss-driven optimization achieved high color fidelity and structural accuracy, resulting in smoother transitions and superior print quality.

\subsection{3D Gaussian Splatting}

In the first attempt using 3D Gaussian Splatting, we observed that the conversion from 2D Gaussian splats to 3D ellipsoidal volumes introduced notable discrepancies in both color fidelity and structural coherence. The continuous-to-discrete color mapping, while effective, led to color inconsistencies, particularly in gradient regions where smooth transitions were essential. Additionally, the intermediate slicing and color matching steps introduced artifacts due to interpolation errors, which slightly degraded the print quality. Overall, this approach underscored the challenges in adapting digital visualization techniques, such as 3DGS, for physical 3D printing.

\subsection{Poxel}

The second attempt, utilizing our Poxel end-to-end model, demonstrated marked improvements in print fidelity and color accuracy. By producing a direct CMYKWCl voxel grid tailored to the printer's capabilities, the Poxel model circumvented the need for conversion steps, reducing artifacts and preserving structural details. The use of anisotropic voxels enabled better resolution control, particularly along the model's height axis, resulting in enhanced print quality and smoother transitions across color gradients. Compared to 3D Gaussian Splatting, the Poxel model provided superior alignment with the printer's color gamut, ensuring that printed colors closely matched their digital counterparts. These results confirm that Poxel's approach is more suitable for high-fidelity, full-color 3D printing than traditional visualization techniques, establishing a streamlined workflow from digital model to physical print.

\section{Future Work}

Future improvements to Poxel will focus on increasing voxel resolution and refining anisotropic structures to enhance fidelity in complex geometries. Optimized voxel placement and distribution will allow smoother surfaces and color transitions without heavy computational costs.

Adaptive color discretization, potentially aided by machine learning, could further improve color fidelity across varied textures. Expanding compatibility to additional 3D printing platforms with flexible profiles would establish Poxel as a versatile tool.

Finally, enhancing computational efficiency through streamlined CUDA processing and distributed computing will enable Poxel to handle high-resolution, full-color models at industrial scale.

\bibliographystyle{IEEEtran}
\bibliography{IEEEabrv,references}

\end{document}